%% file: main.tex
\documentclass[sigconf]{acmart}
\usepackage{multirow, makecell}
\usepackage{listings}
\usepackage[export]{adjustbox}
\usepackage{subfigure}
\usepackage[font=small,skip=0pt]{caption}

\def \ACRONYM {ShareAware}

\AtBeginDocument{%
  \providecommand\BibTeX{{%
    \normalfont B\kern-0.5em{\scshape i\kern-0.25em b}\kern-0.8em\TeX}}}

\copyrightyear{2020}
\acmYear{2020}
\setcopyright{acmcopyright}\acmConference[MM '20]{Proceedings of the 28th ACM International Conference on Multimedia}{October 12--16, 2020}{Seattle, WA, USA}
\acmBooktitle{Proceedings of the 28th ACM International Conference on Multimedia (MM '20), October 12--16, 2020, Seattle, WA, USA}
\acmPrice{15.00}
\acmDOI{10.1145/3394171.3414692}
\acmISBN{978-1-4503-7988-5/20/10}

\begin{document}
\fancyhead{}


\title{Helping Users Tackle Algorithmic Threats on Social Media:\\
A Multimedia Research Agenda}
\author{Christian von der Weth \quad Ashraf Abdul \quad Shaojing Fan \quad Mohan Kankanhalli}
\affiliation{%
  \institution{School of Computing, National University of Singapore}
}
\email{{chris|ashraf|fansj|mohan}@comp.nus.edu.sg}

\renewcommand{\shortauthors}{C. von der Weth, et al.}

\input{sections/abstract}



\maketitle

\input{sections/introduction}
\input{sections/threats}
\input{sections/nudging}
\input{sections/shareaware}

\input{sections/discussion}

\input{sections/conclusions}

\bibliographystyle{ACM-Reference-Format}
\bibliography{bibliography}

\end{document}

%% file: sections/abstract.tex
\begin{abstract}
Participation on social media platforms has many benefits but also poses substantial threats. Users often face an unintended loss of privacy, are bombarded with mis-/disinformation, or are trapped in filter bubbles due to over-personalized content. These threats are further exacerbated by the rise of hidden AI-driven algorithms working behind the scenes to shape users’ thoughts, attitudes, and behaviour. We investigate how multimedia researchers can help tackle these problems to level the playing field for social media users. We perform a comprehensive survey of algorithmic threats on social media and use it as a lens to set a challenging but important research agenda for effective and real-time user nudging. We further implement a conceptual prototype and evaluate it with experts to supplement our research agenda. This paper calls for solutions that combat the algorithmic threats on social media by utilizing machine learning and multimedia content analysis techniques but in a transparent manner and for the benefit of the users.
\end{abstract}

%% file: sections/introduction.tex
\section{Introduction}
\label{sec:introduction}

As platforms for socializing but also as source of information, social media has become one of the most popular services on the Web. However, the use of social media also poses substantial threats to its users. The most prominent threats include a loss of privacy, mis-/disinformation (e.g., ``fake news'', rumors, hoaxes), and ``over-personalized'' content resulting in so-called filter bubbles and echo chambers. While these threats are not new, the risk has significantly increased with the advances of modern machine learning algorithms. Hidden from the average user's eyes, social media platform providers deploy such algorithms to maximize user engagement through personalized content in order to increase ad revenue. These algorithms analyze users' content, profile users, filter and rank content shown to users.
Algorithmic content analysis is also performed by institutions such as banks, insurance companies, businesses and governments. The risk is that institutions attempt to instrumentalize social media with the goal to monitor or even intervene in users' lives~\cite{Guerses2013TwoTalesOfPrivacy}. Lastly, algorithms are getting better in mimicking users. So-called social bots~\cite{Ferrara2016TheRiseOfSocialBots} are programs that operate social media accounts to post and share unverified or fake content. This includes that modern machine learning algorithms can be used to modify or even fabricate content (e.g., \textit{deep fakes}~\cite{Kietzmann2020DeepfakesTrickOrTreat}).

Assessing the risks of these threats is arbitrarily difficult.
Firstly, the average social media user is not aware of or vastly underestimates the power of state-of-the-art machine learning algorithms. Without transparency, users do not know what personal attributes are collected, or what content is shown -- or not shown -- and why. 
\textbf{The lack of awareness and knowledge about machine learning algorithms on the users' part creates an information asymmetry} that makes it difficult for users to effectively and critically evaluate their social media use. 
Secondly, the negative consequences of users' social media behaviour are usually not obvious or immediate.
Concerns such as dynamic pricing or the denial of goods and services (e.g., China's \textit{Social Credit Score}~\cite{Liang2018ConstructingADataDrivenSociety}) are typically the result of a long posting and sharing history. The formation of filter bubbles and their effects on users' views is often a slow process. \textbf{This lack of an immediate connection between users' behavior and negative consequences prohibits an intrinsic incentive for users to change their behavior.}
Lastly, even users who are aware of applied algorithms and negative consequences are in a disadvantaged position. Compared to users, algorithms deployed by data holders have access to a much larger pool of information and to virtually unlimited computing capacities to analyze this information. \textbf{This lack of power leaves users defenseless against the algorithmic threats on social media.}


In this paper, we call for solutions to level the playing field -- that is, to counter the \textit{information asymmetry} that data holders have over the average users. To directly rival the algorithmic threats, we argue for utilizing machine learning and data mining techniques (e.g., multimedia and multimodal content analysis, image and video forensics) but in a transparent manner and for the benefit of the users. The goal is to help users to better quantify the risk of threats, and thus to enable them to make more well-informed decisions. Examples include warnings that a post contains sensitive information before submitting, informing that an image or video has been tampered with, suggesting alternative content from more credible sources, etc. While algorithms towards (some of) these tasks exists (cf. Section~\ref{sec:threats}), we argue that their applicability in our target context of supporting social media users is limited. For interventions (e.g., notices, warnings, suggestions) to be most effective in guiding users behavior, the interventions need to be content-aware, personalized, in-situ, but also trusted and minimally intrusive.

While covering a wide range of research questions from different fields, we believe that the multimedia research community will play an integral part in this endeavour. To this end, we propose a research agenda highlighting open research questions and challenges with focus on multimedia content analysis and content generation. We organize relevant research questions with respect to three core tasks: 
(1) \textit{Risk assessment} addresses the questions of When and Why social media users should be informed about potential threats. 
(2) \textit{Visualization and explanation} techniques need to convert the outcome from machine learning algorithms (e.g., labels, probabilities) to convey potential risks in a comprehensible and relatable manner. 
(3) \textit{Sanitization and recommendations} aim to help users minimize risks either through sanitizing their own content (e.g., obfuscation of sensitive parts in images and videos) or the recommendation of alternative content (e.g., from trusted and unbiased sources).
Lastly, given the user-oriented nature of this research, we also highlight the challenges of evaluating the efficacy of interventions in terms of their influence on social media users' decisions and behavior. To summarize, we make the following contributions:
\begin{itemize}
    \item We present the first comprehensive survey on the threats posed by machine learning algorithms on social media. 
    \item We propose a novel data-driven scheme for user-centric interventions to help social media users to counteract the algorithmic threats on social media.
    \item We raise open research questions to tackle the \emph{information asymmetry} between users and data holders due to an imbalanced access to data and machine learning algorithms. 
\end{itemize}
Sections~\ref{sec:threats} and~\ref{sec:nudging} cover our main contributions, forming the main parts of this paper. We complement these contributions by presenting our current proof-of-concept implementation (Section~\ref{sec:shareaware}) and outlining related research questions and challenges that are equally important but beyond our main research agenda (Section~\ref{sec:discussion}).


%% file: sections/threats.tex
\section{Threats \& Countermeasures}
\label{sec:threats}
Arguably, the most prominent threats of social media are loss of privacy, fake news, and filter bubbles or echo chambers. In this section, we review automated methods that facilitate each threat, and outline existing countermeasures together with their limitations. We use the recent global discourse surrounding the COVID-19 pandemic to help illustrate the relevance of these three threats.

Note that social media has also been associated with a wide range of other threats such as online harassment through bullying, doxing, public shaming, and Internet vigilantism~\cite{Blackwell2017ClassificationAndItsConsequencesforOnlineHarassment}. Social media use has also been shown to be highly addictive, often caused by the fear of missing out (FOMO)~\cite{Blackwell2017ExtraversionNeuroticismAttachmentFOMO}. Being constantly up-to-date with other's lives often leads to social comparison, potentially resulting in feelings of jealousy or envy, which in turn can have negative effects on users' self-esteem, self-confidence, and self-worth~\cite{Vogel2014SocialComparison}. However, these types of threats are generally not directly caused by algorithms and are thus beyond the scope of this work.


\input{sections/threats-privacy}
\input{sections/threats-fake-news}

\input{sections/threats-behavior}

%% file: sections/threats-privacy.tex
\subsection{Loss of Privacy}
\label{sec:threats-privacy}
Each interaction on social media adds to a user's digital footprint, painting a comprehensive picture about the user's real-world identity and personal attributes. 
From a privacy perspective, identity and personal attributes are considered highly sensitive information (e.g., age, gender, location, contacts, health, political views) which many users would not reveal beyond their trusted social circles. However, with the average user being unaware of the capabilities of data holders and with often no immediate consequences, users cannot assess their privacy risks from their social media use. During the COVID-19 outbreak, many infected users shared their conditions online. Apart from reported consequences such as harassment -- e.g., being blamed for introducing COVID-19 into a community -- users might also inadvertently disclose future health issues (with the long-term effects of the infection currently unknown).

\begin{table}[!t]
\caption{Selected approaches for the extraction and inference of privacy-sensitive information from multimedia content.}
\begin{center}
\begin{tabular}{|m{1.2cm}|l|m{3.4cm}|}
    \hline
    \multirow{2}{*}{identity} & biometrics &  \cite{rowley1998neural} \cite{deng2019arcface} \cite{scherhag2019face} \cite{porter1989voice} \cite{schuster2017method} \\
    \cline{2-3}                              
                              & soft biometrics  &  \cite{bulan2017segmentation} \cite{chang2004automatic} \cite{javidi1994optical} \cite{sekou2006credit} \cite{gafurov2007survey} \\
    \hline    
    \multirow{4}{*}{location} & \makecell[l]{user location\\(stable)}  &         \cite{Han2012GeolocationPredictionInSocialMediaData} \cite{Cheng2010YouAreWhereYouTweet} \cite{Mahmud2012WhereIsThisTweetFrom} \cite{Ryoo2014InferringTwitterUserLocations} \cite{Yamaguchi2014OnlineUserLocationInference}
    \cite{Rahimi2015TwitterUserGeolocationUsingUnifiedTextAndNetwork} 
    \cite{Li2012TowardsSocialUserProfiling} \cite{Miura2017UnifyingTextMetadataAndUserNetworkRepresentations} \cite{Rahimi2017NeuralModelForUserGeolocation}    \cite{Do2018TwitterUserGeolocationUsingDeepMultiviewLearning} \\
    \cline{2-3}                              
                              & location in content & \cite{Gelernter2013AnAlgorithmForLocalGeoparsingOfMicrotext} \cite{Lee2014WhenTwitterMeetsFoursquare}  \cite{Li2014FineGrainedLocationExtractionFromTweets} \cite{Hoang2018LocationExtractionFromTweets} \cite{Ritter2011NamedEntityRecognitionInTweets} \cite{li2008modeling} \cite{johnson2002consistent} \cite{li2009landmark}\\
    \cline{2-3}
                              & mobility pattern  & \cite{Hasan2013UnderstandingUrbanHumanActivity} \cite{Gabrielli2014FromTweetsToSemanticTrajectories} \cite{Yuan2017PRED} \cite{Yin2011DiversifiedTrajectoryPatternRanking} \cite{Beiro2016PredictingHumanMobility}\\
    \hline
    \multirow{3}{*}{\makecell[l]{demo-\\graphics}} & gender/age  & \cite{Levi2015AgeGenderClassification} \cite{Sap2014DevelopingAgeAndGenderPredictiveLexica} \cite{You2014GenderPredictionUsingImagesPosted} \cite{Ciot2013GenderInferenceOfTwitterUsersInNonEnglishContexts} \cite{Peersman2011PredictingAgeAndGender} \cite{Burger2011DiscriminatingGenderOnTwitter} \cite{Guimaraes2017AgeGroupsClassificationInSocialNetwork} \cite{Schwartz2013PersonalityGenderAndAge} \cite{Nguyen2013HowOldDoYouThinkIAm} \cite{Nguyen2013TweetGenie} \\
    \cline{2-3}
                                & \makecell[l]{ethnicity/nationality} & \cite{Pietro2018UserLevelRaceAndEthnicityPredictors} \cite{Hofstra2018PredictingEthnicityWithFirstNamesInOSN} \cite{Lee2017NameNationalityClassificationwithRNN} \cite{Ye2017NationalityClassificationUsingNameEmbeddings} \cite{Chang2010ePluribusEthnicityOnSocialNetworks} \\
    \cline{2-3}                              
                                & income/occupation & \cite{Huang2015MultiSourceIntegrationFrameworkForUserOccupationInference} \cite{Matz2019PredictingIndividualLevelIncome} \cite{Hasanuzzaman2017TemporalOrientationOfTweetsForPredictingIncome} \cite{Pietro2015StudyingUserIncome}\\
    \hline
    \multirow{3}{*}{\makecell[l]{perso-\\nality}} & traits  & \cite{Azucar2018PredictingTheBig5PersonalityTraits} \cite{Farnadi2016ComputationalPersonalityRecognition} \cite{Skowron2016FusingSocialMediaCuesPersonalityPrediction} \cite{Leqi2016AnalyzingPersonalityThroughSocialMediaProfilePicture} \cite{Park2015AutomaticPersonalityAssessment} \cite{Ferwerda2015PredictingPersonalityTraitsWithInstagramPictures} \cite{Hughes2012ATaleOfTwoSites} \cite{Sumner2012PredictingDarkTriadPersonalityTraits} \cite{Adali2012PredictingPersonalityWithSocialBehavior} \cite{Golbeck2011PredictingPersonality} \\
    \cline{2-3}
                                & emotion \& mood & \cite{Bargal2016EmotionRecognitionInTheWildFromVideosUsingImages} \cite{Nojavanasghari2016EmoReact} 
                                \cite{liong2016spontaneous} \cite{borza2017high} \cite{majumder2019dialoguernn}\\
    \cline{2-3}
                                & sexual orientation & \cite{Wang2018DNNSexualOrientation} \cite{Jernigan2009Gaydar} \\
    \hline
    \multirow{3}{*}{health} & (social) well-being  & \cite{Chen2017BuildingProfileOfSubjectiveWellBeing} \cite{Schwartz2016PredictingIndividualWellBeing} \cite{McIver2015CharacterizingSleepIssuesUsingTwitter} \cite{Sueki2015TheAssociationOfSuicideRelatedTwitterUseWithSuicidalBehaviour} \cite{Burke2010SocialNetworkActivityAndSocialWellBeing} \cite{doty2013fearful} \cite{giannakakis2017stress}\\
    \cline{2-3}
                              & mental health & 
                             \cite{Bishay2019SchiNet} \cite{Cook2019TowardsAutomaticScreeningOfTypicalAndAtypicalBehaviorsInChildrenWithAutism} \cite{Guntuku2017DetectingDepressionAndMentalIllness} \cite{Benton2017MultitaskLearningForMentalHealthConditions} \cite{Chancellor2016QuantifyingAndPredictingMentalIllnessSeverity} \cite{DeChoudhury2016DiscoveringShiftsToSuicidalIdeation} \cite{DeChoudhury2013PredictingDepressionViaSocialMedia} \cite{Rajagopalan2013SelfStimulatoryBehavioursInTheWildForAutismDiagnosis} \cite{Wald2012UsingTwitterContentToPredictPsychopathy} \cite{Rehg2014BehavioralImagingAndAutism} \cite{wang2016revealing} \\
    \cline{2-3}                              
                              & physical health & \cite{Merchant2019EvaluatingThePredictabilityOfMedicalConditions} \cite{Kocabey2017FaceToBMI} \cite{Weber2016CrowdsourcingHealthLabels} \cite{Wen2013ComputationalApproachToBodyMassIndexPredictionFromFaceImages} \\                                
    \hline    
    \multirow{3}{*}{\makecell[l]{relation-\\ships}} & tie strength & \cite{Gilbert2009PredictingTieStrengthWithSocialMedia} \cite{Panovich2012TieStrengthIInQaAndSns} \cite{Gupte2012MeasuringTieStrength} \cite{Burke2014GrowingCloserOnFacebook} \\
    \cline{2-3}                              
                                & tie role & \cite{Zhao2013InferringSocialRolesAndStatusesInSocialNetworks} \cite{Zhang2018FromFacialExpressionRecognitionToInterpersonalRelationPrediction} \cite{Guo2019SocialRelationshipRecognitionBasedOnAHybridDeepNeuralNetwork} \cite{Li2017DualGlanceModelForDecipheringSocialRelationships} \cite{Wang2018DeepReasoningWithKnowledgeGraphForSocialRelationshipUnderstanding} \\
    \cline{2-3}                              
                                & community & \cite{Hofstra2017SourcesOfSegregationInSocialNetworks} 
                                \cite{Bedi2016CommunityDetectionInSocialNetworks} \cite{Du2007CommunityDetectionInLargeScaleSn} \cite{Wang2015CommunityDetectionInSn}\\                           
    \hline
    \multirow{2}{*}{\makecell[l]{beliefs,\\opinions,\\lifestyle}} & religion  & \cite{Nguyen2014OnPredictingReligionLabelsInMicrobloggingNetworks} \cite{Chen2014USReligiousLandscapeOnTwitter} \cite{Yaden2018TheLanguageOfReligiousAffiliation} \\
    \cline{2-3}
                                & political leaning & \cite{Chang2017PredictingPoliticalAffiliationOfPostsOnFacebook} \cite{Pietro2017BeyondBinaryLabelsPoliticalIdeologyPrediction} \cite{Wong2016QuantifyingPoliticalLeaningFromTweets} \cite{Colleoni2014EchoChamberOrPublicSphere} \cite{Pla2014PoliticalTendencyIdentificationInTwitter} \cite{Cohen2013ClassifyingPoliticalOrientationOnTwitter} \cite{Makazhanov2013PredictingPoliticalPreferenceOfTwitterUsers} \cite{Conover2011PredictingThePoliticalAlignmentOfTwitterUsers} \cite{Golbeck2011ComputingPoliticalPreferenceAmongTwitterFollowers} \\
                                \cline{2-3}
                                & habits & \cite{DeChoudhury2016CharacterizingDietaryChoicesNutrition} \cite{Pagoto2014TweetingItOffWeightLossAttempt} \cite{Myslin2013UsingTwitterToExamineSmokingBehaviorAndPerceptionsOfEmergingTobaccoProducts} \cite{Prochaska2012TwitterQuitter} \\

    \hline        
\end{tabular}
\end{center}
\label{tab:overview-privacy-threats}
\end{table}
\textbf{Algorithmic threats.}
Given its value, a plethora of algorithms have been designed to extract or infer personal information from multimedia content about virtually all aspects of a user's self: identity, personality, health, location, beliefs, relationships, etc. Table~\ref{tab:overview-privacy-threats} presents and overview of the main types of personal information with a selection of related works. While not all methods have been proposed explicitly with social media content in mind (e.g., face detection \cite{hjelmaas2001face,hsu2002face} and action recognition \cite{varol2017long,crasto2019mars} algorithms which are generally used for security surveillance) they can in principle be applied to content posted and shared on social media. Space constraints prohibit a more detailed discussion, but the key message is that algorithms that put user's privacy at risk are omnipresent.

\textbf{Existing countermeasures.}
Data and information privacy has been addressed from technological, legal and societal perspectives. With our focus on using technology to protect users' privacy in social media, we categorize existing approaches as follows:

\textit{(1) Policy comprehension} aims to help users understand their current privacy settings. A basic approach is to show users their profile and content from the viewpoint of other users or the public~\cite{Lipford2008AudienceView,Anwar2012AVisualizationTool}. More general solutions propose user interface design elements (color-coding schemes, charts, graphs) to visualize which content can be viewed by others~\cite{Paul2012C4PS,Mazzia2012PViz,DeWolf2015ThePromiseOfAudienceTransparancy,Stern2014ImprovingPrivacySettingsWheel,Hu2011DetectingAndResolvingPrivacyConflicts}. 

\textit{(2) Policy recommendation} techniques suggest privacy settings for newly shared data items. Social context-based methods assign the same privacy settings to similar contacts. Basic approaches consider only the neighbor network of a user~\cite{AduOppong2008SocialCircles,Danezis2009InferringPrivacyPolicies}), while more advanced methods also utilize profile information~\cite{Fang2010PrivacyWizard,Jones2010FeasibilityOfClusteringForPolicy,McAuley2012LearningToDiscoverSocialCircles,Squicciarini2012AutomaticSocialGroupOrganization,Amershi2012ReGroup,Misra2016IMPROVE}). Content-based policy recommendations assign privacy settings based on the similarity of data items. First works used (semi-)structured content such as tags or labels ~\cite{Vyas2009TowrdsAutomaticPrivacyManagement,Yeung2009ProvidingAccessControl,Klemperer2012TagYouCanSeeIt,Ravichandran2009CapturingSocialNetworkingPrivacy}. With the advances in machine learning, solutions have been proposed that directly analyze unstructured content with emphasis on images~\cite{Squicciarini2017TowardAutomatedOnlinePhotoPrivacy,Zerr2012PrivacyAwareImageClassification,DeChoudhury2009ConnectingContentToCommunity,Squicciarini2015PrivacyPolicyInference} as well as textual content~\cite{CaliskanIslam2014PrivacyDetective,Naini2015AnalyzingPredictingPrivacySettings,Chen2018PrivacyPredictionModel}. 
State-of-the art deep learning models allow for end-to-end learning and yield superior results~\cite{Yu2017iPrivacy,Yu2018LeveragingContentSensitiveness,SpyromitrosXioufis2016PersonalizedPrivacyAwareImage}.
More recent policy recommendations aim to resolve conflicts in case of different privacy preferences for co-owned objects~\cite{Such2018MultipartyPrivacy}.

\textit{(3) Privacy nudging}~\cite{Acquisti2017NudgesForPrivacyAndSecurity,Wang2014FieldTrialPrivacyNudges} introduces design elements or makes small changes to the user interface to remind users of potential consequences before posting content and to rethink their decisions: timer nudges delay the submission of a post; sentiment nudges warn users that their post might be viewed as negative; audience nudges show a random subset of contacts to remind users of who will be able to view a post.

Most of the proposed solutions focus on users' privacy amongst their social circles -- that is, privacy is preserved if a user's content can only be viewed by others in line with the user's intention. This assumes that social media platform providers are trusted to adhere to all privacy settings but also that no threats are coming from the platform providers and any data holders with access to users' content. Only privacy nudging does not explicitly rely on trust in data holders. However, current privacy nudges are content-agnostic. They neither provide detailed information why the content might be too sensitive to share nor offer suggestions to users on how to edit or modify content submissions to lower their level of sensitivity.

%% file: sections/threats-fake-news.tex
\subsection{Fake News}
\label{sec:threats-fake-news}
Mis- and disinformation has long been used to shape people's thoughts and behavior, but social media has significantly amplified its adverse effects. Fake news often leverages users' cognitive biases (e.g., confirmation bias, familiarity bias, anchoring bias), making it more likely for users to fall for it~\cite{Pennycook2019WhoFallsForFakeNews}. Fake news is typically also novel, controversial, emotionally charged, or partisan, making it more ``interesting'' and hence more likely to be shared~\cite{Vosoughi2018TheSpreadOfTrueAndFalseNewsOnline}. During the COVID-19 crisis, misleading claims about infection statistics have resulted in delayed or ignored counteractions (e.g., social distancing measures). False conspiracy theories about the causes for the disease have, for example, resulted in destroying 5G communication towers. More tragically, fake news about supposed cures have already cost the lives of several hundred people.

\textbf{Algorithmic threats.}
The most popular algorithmic threat for spreading fake news is the use of bots to spread mis-/disinformation. Particularly bots created with malicious intent aim to mimic humanlike behavior to avoid detection efforts and trick genuine users into following them. This enables the bots to capture large audiences, making it easier to spread fake news. Mimicking humanlike behavior may include varying sharing frequency and schedule, or periodically updating profile information~\cite{Ferrara2016TheRiseOfSocialBots,Varol2017OnlineHumanBotInteractions}. Apart from better ``blending in'', sophisticated bots also coordinate attacks through the synchronization of whole bot networks~\cite{Grimme2018ChangingPerspectives}. Recent advances in machine learning also allow for the automated doctoring or fabrication of content. This includes the manipulation of multimedia content such as text swapping~\cite{Kietzmann2020DeepfakesTrickOrTreat} or image splicing~\cite{Zhang2020AnOverviewOfOnlineFakeNews}. When coupled with, algorithms  used to detect ``infectious'' multimedia i.e., content that is most likely to go viral~\cite{Tatar2014SurveyOnPredictingPopularity,Deza2015UnderstandingImageVirality,Han2017PredictingPopularAndViralImage}, they can be used to predict the effectiveness of fabrication, e.g., for the generation of clickbait headlines~\cite{Shu2018DeepHeadlineGeneration}. Finally, fake content can also be generated using Generative Adversarial Networks (GANs), a deep learning model for the automated generation of (almost) natural text, images or videos. The most popular example audio-visual content are so-called ``deep fakes''~\cite{Kietzmann2020DeepfakesTrickOrTreat}: videos that show, e.g., a politician making a statement that never occurred. For textual content, the Generative Pre-trained Transformer 3 (GPT-3)~\cite{Brown2020LanguageModelsAreFewShotLearners} represents the current state of the art of generating humanlike text.

\textbf{Existing countermeasures.} 
The effects of fake news saw many countries introduce laws imposing fines for its publication~\cite{Roudik2019InitiativesToCounterFakeNews}. However, the vague nature of fake news makes it very difficult to put it into a legal framework~\cite{Klein2017FakeNewsLegalPerspective} and raises concerns regarding censorship and misuse~\cite{Vosoughi2018TheSpreadOfTrueAndFalseNewsOnline}. Other efforts include public information campaigns or new school curricula that aim to improve critical thinking skills and media literacy. However, these are either one-time or long-term efforts with uncertain outcomes~\cite{Lazer2018TheScienceOfFakeNews}. From a technological perspective, a plethora of data-driven methods have been proposed for the identification of social bots, automated credibility assessment, and fact-checking~\cite{Shu2017FakeNewsDetectionOnSocialMediaDataMiningPerspective}. 
Most methods to identify social bots use supervised machine learning by leveraging on the user, content, social network, temporal features, etc. (e.g.,~\cite{Yang2020ScalableAndGeneralizableSocialBotDetectionThroughDataSelection,Kudugunta2018DeepNeuralNetworksForBotDetection}). Unsupervised methods aim to detect social bots by finding accounts that share strong similarities with respect to social network and (coordinated) posting/sharing behavior (e.g.,~\cite{Chavoshi2016DeBot,Mazza2019RTbust}). Fact-checking is a second corner stone to counter fake news. However, manual fact-checking -- done by websites such as Snopes or Politifact, or dedicated staff of social media platform providers -- scales poorly with the amount of online information. Various solutions for automated fact-checking have been proposed (e.g., \cite{Karadzhov2017FullyAutomatedFactCheckingUsingExternalSources,Hassan2017TowardAutomatedFactChecking}). However, fully automated fact-checking systems are far from mature and most real-world solutions take a hybrid approach~\cite{Graves2018UnderstandingThePromiseAndLimitsOfAutomatedFactChecking}. 

Existing technological solutions to combat fake news focus on the ``bad guys'' and do not address the impact of the average user on its success. However, Vosoughi et al.~\cite{Vosoughi2018TheSpreadOfTrueAndFalseNewsOnline} have shown that false information spreads fast and wide even without the contributions by social bots. To evaluate the effects of user nudging, Nekmat~\cite{Nekmat2020NudgeEffectOfFactCheckAlerts} conducted a series of user surveys to evaluate the effectiveness of fact-check alerts (e.g., the reputation of a news source). The results show that such alerts trigger users' skepticism, thus lowering the likelihood of sharing information from questionable sources. Similarly, Yaqub et al.~\cite{Yaqub2020EffectsOfCredibilityIndicators} carried out an online study to investigate the effects of different credibility indicators (fact checkers, mainstream media, public opinion, AI) on sharing. The effects differ not only across indicators -- with fact checkers having the most effect -- but also across demographics, social media use, and political leanings.

%% file: sections/threats-behavior.tex
\subsection{Filter Bubble \& Echo Chambers}
\label{sec:threats-groupthink}
Personalized content is one of the main approaches social media platform providers use to maximize user engagement: users are more likely to consume content that is aligned with their beliefs, opinions and attitudes. This selective exposure has led to the rise of phenomena such as echo chambers and filter bubbles~\cite{Bakshy2015ExposureToIdeologicallyDiverseNewsAndOpinionOnFacebook,Barbera2015TweetingFromLeftToRight,Bozdag2015BreakingTheFilterBubble}. The negative effects are similar to the ones of fake news. Here, not (necessarily) false but one-sided or skewed information leads to uninformed decisions particularly in politics~\cite{Beam2018FacebookNewsAndDePolarization,Dylko2017TheDarkSideOfTechnology}. Filter bubbles and echo chambers also amplify the impact of fake news since ``trapped'' users are less likely be confronted with facts or different opinions; as was also the case for COVID-19~\cite{Cinelli2020Covid19SocialMedia}.

\textbf{Algorithmic threats.}
Maximizing user engagement through customized content is closely related to the threats of privacy loss and fake news -- that is, the utilization of personal information and the emphasis on viral content. A such, most of the algorithmic threats also apply here. An additional class of algorithms that often result in ``over-customization'' are recommender systems based on collaborative filtering~\cite{Bobadilla2013RecommenderSystemsSyrvey,Bellogin2013ComparativeStudyOfHeterogeneousItemRecommendationsInSocialSystems,Portugal2017TheUseOfMachineLearningAlgorithmsInRecommenderSystems}. They are fundamental building blocks of many online services such as online shopping portals, music or video streaming sites, product and service review sites, online news sites, etc. Recommender systems aim to predict users' preferences and recommend items (products, songs, movies, articles, etc.) that users are likely to find interesting. Social network and social media platforms in particular expand on this approach by also incorporating the information about a user's connections 
into the recommendation process~\cite{Tang2013SocialRecommendationAReview,Zhou2012StateOfTheArtInPersonalizedRecommenderSystems}. Recommender systems continue to be a very active field of research~\cite{Zhang2019DeepLearningBasedRecommenderSystems,Batmaz2018ReviewOnDeepLearningForRecommenderSystems}.

\textbf{Existing countermeasures}.
Compared to fake news, filter bubble and echo chambers are only a side effect of personalized content~\cite{Burbach2019BubbleTrouble}. 
While recommendation algorithms for incorporating diversity have been proposed (e.g.,~\cite{Tsai2018BeyondTheRankedLists,Lunardi2019RepresentingTheFilterBubble}), they are generally not applied by platform providers since they counter the goal of maximizing user engagement. As a result, most efforts to combat filter bubbles aim to raise users' awareness and give them more control~\cite{Bozdag2015BreakingTheFilterBubble}. From an end user perspective, multiple solutions propose browser extensions that analyze users' information consumption and provide information about their reading or searching behavior and biases; e.g.,~\cite{Munson2013EncouragingReadingOfDiversePoliticalViewpoints,Xing2014ExposingInconsistentWebSearchResultsWithBobbleFaridani2010OpinionSpace}. 
All these approaches can be categorized as nudging by making users be aware of their biases.
Despite the numerous nudging measures proposed, their application and acceptance among social media users remain low \cite{lehner2016nudging,weinmann2016digital}. This may be because of several reasons. Firstly, as an emerging new technology, digital nudging is not as established as expert nudging, thus may lack users' trust in the first place~\cite{weinmann2016digital}. Secondly, many people use social media due to its convenience and pleasance, whereas nudging requires extra efforts and attention. Finally, psychologists have found that nudging might only have effects on things people are truly aware of and care about~\cite{french2011nudging}. 

%% file: sections/nudging.tex
\section{User Nudging in Social Media}
\label{sec:nudging}
To address the information asymmetry between users and data holders,
we motivate automated and data-driven user nudging as a form of educational intervention.
In a nutshell, nudges aim to inform, warn or guide users towards a more responsible behavior on social media to minimize the risk of potential threats such as the loss of privacy or the influence of fake news. In this section, we propose a research agenda for the design and application of effective user nudges in social media. 

\input{sections/nudging-design-goals}
\input{sections/nudging-tasks}

\input{sections/nudging-engine}

\input{sections/nudging-evaluation}

%% file: sections/nudging-design-goals.tex
\subsection{Design Goals}
\label{sec:nudging-design-goals}
Effective nudges should be helpful to users without being annoying. Presenting users too often with bad or unnecessary information or warnings may result in users ignoring nudges. We formulate the following design goals for effective nudging:

\textbf{(1) Content-aware.} 
Warning messages should only be displayed if necessary, i.e., if a potential risk has been identified. For example, an image showing a generic landscape is generally less harmful compared to an image containing nudity. Thus, the latter would more likely trigger a nudge. Similarly, only content from biased or untrusted sources, or content that show signs of being tampered with should result in the display of warnings. Effective nudges need to be tailored to content such as articles or posts being shared.

\textbf{(2) User-dependent.}
The need and the instance of a nudge should depend on the individual users, with the same content potentially triggering different or no nudges for different users. For example, a doctor revealing her location in a hospital is arguably less sensitive compared to other users. Similarly, a user who is consciously and purposefully reading articles from different news sites, does not need a warning about each site's biases.

\textbf{(3) Self-learning.}
As consequence of both content- and user-dependency, a nudging engine should adapt to a user's social media use. To be in line with the idea of soft paternalism, users should be in control of nudges through manual or (semi-)automated personalization. This personalization might be done through explicit feedback by the user or through implicit feedback derived from the user's past behavior (e.g., the ignoring of certain nudges). 

\textbf{(4) Proactive.} 
While deleting content shortly after posting might stop it from being seen by other users, it is arguably still stored on the platform and available for analysis. Assuming untrusted data holders, any interaction on social media has to be considered as permanent. Thus, nudges need to be displayed before any damage might be done, e.g., before privacy-sensitive content is submitted, a fake news or biased article is shared or even read, etc.

\textbf{(5) In-situ.} 
Educational interventions are most effective when given at the right time and the right place. Therefore, nudges should be as tightly integrated into users every-day social media use as possible. Ideally, a false fact is highlighted in a news article, sensitive information is marked within an image, warning messages are displayed near the content, etc. The level of integration depends on the environment, with desktop browsers being more flexible compared to closed mobile apps (cf. Section~\ref{sec:discussion}).

\textbf{(6) Transparent.}
Data-driven user nudging relies in many cases on similar algorithms as potential attackers (e.g., to identify privacy-sensitive information). In contrast to the hidden algorithms of data holders, the process of user nudging therefore needs to be as open and transparent as possible to establish users' trust. Transparency also requires explainability -- that is, users need to be able to comprehend why a nudge has been triggered to better adjust their social media use in the future.


%% file: sections/nudging-tasks.tex
\subsection{Core Tasks}
\label{sec:nudging-tasks}
We argue that an effective nudging engine contains three core components for the tasks of risk assessment, representation and visualization of nudges, and the recommendation or sanitization of content. In the following section, we outline the challenges involved and derive explicit research questions for each task.

\input{sections/nudging-task-identification}

\input{sections/nudging-task-explanation}
\input{sections/nudging-task-sanitization}

%% file: sections/nudging-task-identification.tex
\textbf{Risk assessment} 
refers to the task of identifying the need for nudges in case of, e.g., privacy-sensitive content, tampered content, fakes news or biased sources, clickbait headlines, social bots, etc. As such, risk assessment can leverage on existing methods outlined in Section~\ref{sec:threats}. Note that user nudging therefore relies on the same algorithms used by attackers; we discuss ethical questions and concerns in Section~\ref{sec:discussion}. Despite the availability of such algorithms, their applicability for risk assessment is arguably limited with respect to our outlined design goals. Not always is an image containing nudity privacy-sensitive (e.g., an art exhibition), not every bot has a malicious intent (e.g., weather or traffic bots), not always is a deep fake video shared to deceive but only to entertain users. Effective risk assessment therefore requires a much deeper understanding of content and context. We formulate these challenges with the following research questions:

\begin{itemize}
    \item How can existing countermeasures against threats in social media be utilized for risk assessment towards user nudging?
    \item What are the shortcomings of existing methods that limit their applicability for risk assessment with respect to the design goals (particularly to minimize the number of nudges)?
    \item How to design novel algorithms for effective risk assessment with a deep(er) semantic understanding of the content?
\end{itemize}

%% file: sections/nudging-task-explanation.tex
\textbf{Generation and visualization}
addresses the task of presenting nudges to users. Sophisticated solutions of risk assessment rely on modern machine learning methods that return results beyond the understanding of the average social media user. Firstly, the outcomes of those algorithms are generally not intuitive: class labels, probabilities, scores, weights, bounding boxes, heatmaps, etc., making it difficult for most users to interpret those outcomes. And secondly, the complexity of most methods makes it difficult to comprehend how or why a method returned a certain outcome. Such explanations, however, would greatly improve the trust in outcomes and thus nudges. The need for understanding the inner workings of machine learning methods spurred the research field of eXplainable AI (XAI) to make such models (more) interpretable. However techniques so far are targeted primarily for experts and improving their usability for end users is an active area of research \cite{abdul2018trends,gilpin2018explaining,abdul2020cogam}. Regarding the generation of nudges, we formulate the following research questions:
\begin{itemize}
    \item To what extent are current XAI methods applicable for user nudging in social media?
    \item How to convert outcomes and existing explanations into a more readable and user-friendly format for nudging (e.g., charts, content markup or highlighting, verbalization)?
    \item How to measure the efficacy of nudges along human factors such as plausibility, simplicity, relatability etc. to evaluate the trade-off between these opposing goals?
    \item How to make the generation of nudges customizable to accommodate users' preferences and expertise or knowledge?
\end{itemize}

With solutions for generating nudges available, the last step concerns the questions of how to display nudges. Very few works have investigated the effects of different aspects of nudges in the context of privacy~\cite{schaub2015design,balebako2015impact,gluck2016short} and credibility of news articles~\cite{poston2005effective,lin2016social,zhang2018structured,Nekmat2020NudgeEffectOfFactCheckAlerts}. Based on previous works, we can define four key aspects for visualizing user nudges: (1) timing, i.e., the exact time when a user nudge is presented, (2) location, i.e., the places where a nudge is presented, (3) format, i.e., the media formats in which the information is presented (whether audio and visual information should be combined with textual information, the length of the information), and (4) authorization, i.e., users' control over the nudging information. We formulate the following research questions for these aspects as follows:
\begin{itemize}
    \item Given the different threats in social media, what kind of information would users find most useful (e.g., visualization method, level of detail, auxiliary information)?
    \item How does the \textit{How}, \textit{When} and \textit{Where} of users' control effect the effectiveness of nudges on the behavior of users?
\end{itemize}

%% file: sections/nudging-task-sanitization.tex
\textbf{Recommendation and sanitization} expand on nudges that assess and visualize risks of threats to also include suggestions to lower those risks to further support and educate users. In case of fake news, biased sources or tampered content, such suggestions would include the recommendation of credible and unbiased sources, or links to the original content. Regarding our design goals (here: content-awareness) recommended alternative content must be similar or relevant to the original content. This refers to the fundamental task of measuring multimedia content similarity and related tasks such as reverse image search. However, besides the similarity of content, suitable metrics also need to incorporate new aspects such as a classification of the source (e.g., credibility, biases, intention). For recommending alternative content, we propose the following research questions:
\begin{itemize}
    \item How can existing similarity measures and content linking techniques be applied to suggest alternative content or sources for nudging to lower users' risks?
    \item How can those methods be extended to consider additional aspects beyond raw content similarity?
\end{itemize}

In case of users creating content, a more interesting form of suggestion involves the modification of the content to reduce any risks. This is particularly relevant for privacy risks where already minor modifications may avoid a harmful disclosure. However, quickly and effectively editing images or videos is beyond the skills of most users. Content sanitization, the automated removal of sensitive information from content, is a well-established task in the context of data publishing to facilitate downstream analysis tasks of user information without infringing on their privacy. However, with very few exceptions (e.g., ~\cite{Murugesan2013tPlausability,Shen2019HumanImperceptiblePrivacyProtectionAgainstMachines}), the sanitized content is not intended to be viewed by users. This makes  techniques such as word removal, as well as the cropping, redaction or blurring of images or videos valid approaches. In contrast, content sanitization for social media, where the output is seen by others, must fulfil two requirements: 
\textit{(1) Preservation of integrity}. Any sanitization must preserve the integrity of the content -- that is, sanitized content must read or look natural and organic, and it should not be obvious that the content has been modified.
\textit{(2) Preservation of intention.} The sanitized content should reflect the user's original intention for posting as much as possible to make it a more valid alternative for the user to consider -- formulated as research questions:
\begin{itemize}
    \item What are limitations of existing content sanitization techniques w.r.t their applicability for nudging in social media?
    \item How can the integrity of content be measured to evaluate if a sanitized text, image or video appears natural and organic?
    \item How can a user's intention be estimated to guide the sanitization process towards acceptable alternatives?
    \item How to design, implement and evaluate novel techniques for sanitizing text, images and videos that preserve both content integrity and user intention?
\end{itemize}

%% file: sections/nudging-engine.tex
\subsection{Nudging Engine}

The nudging engine refers to the framework integrating the solutions for the core tasks of risk assessment, generation and visualization, content recommendation and sanitization, as well as additional task for the personalization and configuration for users. 

\textbf{Frontend.} The frontend facilitates two main tasks. Firstly, it displays nudges to the users. For our current prototype, we use a browser extension that directly injects nudges into the website of social media platforms. This includes that the extension can intercept requests to analyze new post before they are submitted; see Section~\ref{sec:shareaware} for more details. And secondly, the frontend has to enable the configuration and personalization of the nudging. To this end, the frontend needs to provide a user interface for manual configuration and providing feedback. To improve transparency, configuration should include privacy settings. For example, a user should be able to select whether a new post is analyzed on its own or in combination with the user's posting history. On the other hand, the frontend should also support automated means to infer users' behavior and preferences. For example, if the same or similar nudges gets repeatedly ignored, the platform may no longer display such nudges in the future. The following research questions summarize the challenges for developing the frontend:
\begin{itemize}
    \item How can nudges be integrated into different environments, mainly desktop browsers and mobile devices?
    \item What means for configuring and providing feedback offer the best benefits and transparency for users?
    \item What kind of data should users provide that best reflect their preferences regarding their consideration of nudges?
\end{itemize}

\textbf{Backend.} The backend features all algorithms for content analysis (risk assessment), content linking (recommendations) and content generation/modification (sanitization). Many of these tasks may rely on external data sources for bot detection, fact-checking, credibility and bias analysis. Depending on users' preferences (see above), the backend will also have to store user content (e.g., users' posting history) and perform behavior analysis to personalize nudges for the individual users. By default, the backend should only keep as much user data as needed -- formulated as research questions:
\begin{itemize}
    \item What are suitable methods to infer user preferences for an automated personalization of nudges?
    \item Where should user data be maintained to optimize the trade-off between user privacy and the effectiveness of nudges?
    \item How to identify and utilize relevant external knowledge to complement user data to further improve nudging?
\end{itemize}

%% file: sections/nudging-evaluation.tex
\subsection{Evaluation}
\label{sec:nudging-evaluation}
User nudging as a form of educational intervention is very subjective with short-term and long-term effects on users' behavior. Few existing works have conducted user studies (e.g., for privacy nudges~\cite{Wang2014FieldTrialPrivacyNudges,Acquisti2017NudgesForPrivacyAndSecurity}) or included user surveys (e.g., for fake news nudges~\cite{Nekmat2020NudgeEffectOfFactCheckAlerts}) in their evaluation. However, evaluating the long-term effectiveness of user nudges on a large scale is an open challenge. While solutions for the core tasks (cf. Section~\ref{sec:nudging-tasks}) can generally be evaluated individually, evaluating the overall performance of the nudging engine is not obvious. We draw from existing efforts towards the evaluation of information advisors. One of the earliest study~\cite{urban1999design} evaluated a trust-based advisor on the Internet, and used users' feedback in interviews as a criteria for the advisor's efficacy. An effective assessment should serve multiple purposes, measure multiple outcomes, and draw from multiple data sources and use multiple methods of measurements~\cite{cuseo2008assessing}. 

Similarly, we propose four dimensions from multiple disciplines to evaluate the efficacy of user nudging, namely influence, trust, usage, and benefits. As the most important dimension, \emph{influence} qualitatively measures if and how user behaviour is influenced by user nudging. The \emph{trust} dimension evaluates the confidence of the user in nudges. Since nudges are also data-driven, it is not obvious why users should trust those algorithms more than those of Facebook, Twitter, and the like. \emph{Usage}, as the most objective dimension, measures the frequency of the use of nudges by users. In contrast, the \emph{benefits} for a user are highly subjective as it requires to evaluate how the user has benefited from nudges, which can mostly be achieved through surveys or interviews. Objective and subjective measures from psychology and economics are needed to evaluate the efficacy of user nudging in a comprehensive manner. We formulate the following research questions:
\begin{itemize}
    \item What are important objective and subjective metrics that quantify the efficacy of user nudging?
    \item How can particularly the long-term effects of nudges on users' social media use be evaluated?
    \item How can the effects of nudges on the threats such as fake news or echo chambers be evaluated on a large scale?
    \item How to evaluate the efficacy of nudging on a community level instead of from an individual user's perspective?
\end{itemize}

%% file: sections/shareaware.tex
\section{\ACRONYM{} Platform}
\label{sec:shareaware}
To make our goals and challenges towards automated user nudging in social media more tangible, this sections presents \ACRONYM{} our early-stage prototype for such a platform. 

\input{sections/shareaware-prototype}

\input{sections/shareaware-evaluation}

%% file: sections/shareaware-prototype.tex
\subsection{Overview to Prototype}
\label{sec:shareaware-prototype}
For a seamless integration of nudges into users' social media use, we implemented the frontend of \ACRONYM{} as a browser extension. This extension intercepts user actions (e.g., posting of new content or sharing of existing content), sends content to the backend for analysis, and displays the results in form of warning messages. The content analysis in the backend is currently limited to basic features to assess the feasibility and challenges of such an approach. In the following examples, we focus on the use case where a user wants to submit a new tweet. If the analysis of a tweet results in nudges, the user can cancel the submission, submit the tweet ``as is'' immediately, or let a countdown run out for an automated submission (similar to a timer nudge~\cite{Wang2014FieldTrialPrivacyNudges}).

\textbf{\ACRONYM{} for privacy.}
\begin{figure}
    \centering
    \includegraphics[width=0.46\textwidth,cfbox=lightgray 0.2pt 1pt]{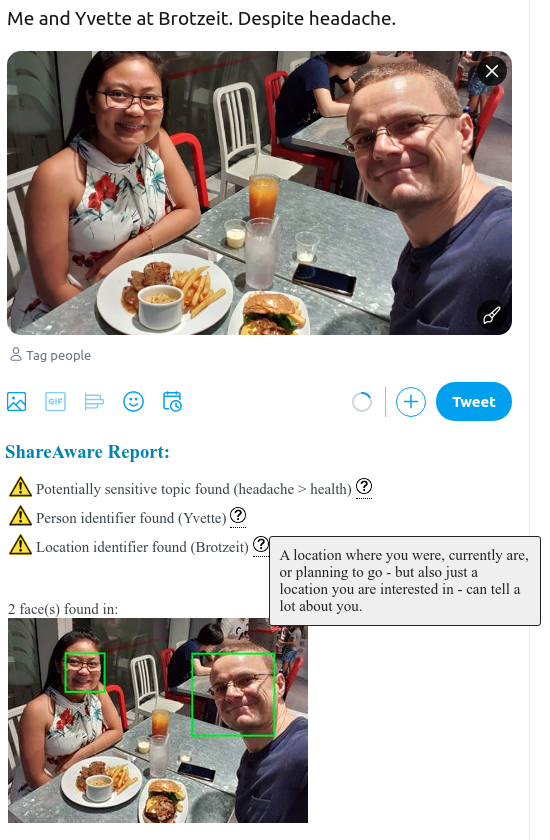}
    \caption{Example of warning messages privacy protection.}
    \label{fig:shareaware-browser-extension-twitter-05-min}
\end{figure}
To identify privacy-sensitive information in text, \ACRONYM{} currently utilizes basic methods such pattern matching (e.g., for identifying phone number, credit card numbers, email addresses), public knowledge graphs such as WordNet~\cite{Miller1995WordNet} and Wikidata~\cite{Vrandecic2014Wikidata} (e.g., to associate ``headache'' with the sensitive topic of health), and Named Entity Recognition (NER) to identify person and location names. The analysis of images is currently limited to the detection of faces. Figure~\ref{fig:shareaware-browser-extension-twitter-05-min} provides an example. More reliable and effective methods to identify privacy-sensitive information will require a much deeper semantic understanding of shared content. A first follow-up step will be to evaluate algorithms outlined in Table~\ref{tab:overview-privacy-threats} for their applicability in \ACRONYM{}.

\begin{figure}
    \centering
    \includegraphics[width=0.46\textwidth,cfbox=lightgray 0.2pt 1pt]{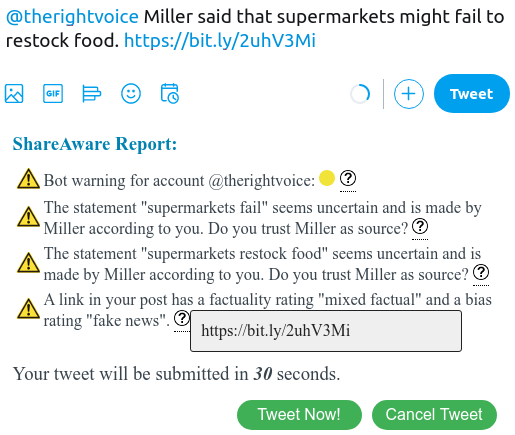}
    \caption{Example of warning messages regarding fake news.}
    \label{fig:shareaware-fake-news-browser-extension-twitter-04}
\end{figure}

\textbf{\ACRONYM{} against fake news.} To slow down the spread of fake news, we display three types of warning messages; see Figure~\ref{fig:shareaware-fake-news-browser-extension-twitter-04}. Firstly, we leverage on the Botometer API~\cite{Yang2020ScalableAndGeneralizableSocialBotDetectionThroughDataSelection} returning a score representing the likelihood that a Twitter account is a bot. We adopt this score but color-code it for visualization. Secondly, we display credibility information for linked content using collected data for 2.7k+ online news sites provided by Media Bias Fact Check (MBFC).\footnote{https://mediabiasfactcheck.com/} MBFC assigns each news site one of six factuality labels and one of nine bias or category labels. We show these labels as part of warning messages. Lastly, we perform a linguistic analysis to identify if a post reflects the opinion of the tweet author or whether the author refers another source making the statement (e.g., \textit{``Miller said that...''}). In case of the latter, we nudge a user accordingly and ask if s/he trust the source. We present \ACRONYM{} for fake news together with an evaluation in a related paper~\cite{vdw2020NudingUsersToSlowDown}.

%% file: sections/shareaware-evaluation.tex
\subsection{Experts Feedback}
\label{sec:shareaware-evaluation}
Our current prototype is still in a conceptual stage. To further assess the validity of the concept as well as understand how and if it can be effectively translated into practice, we perform a think aloud feedback session with 4 experts  (2 Multimedia + 2 Human Computer Interaction researchers). Each session lasted  45-60 minutes and consisted of discussions centred around different scenarios of use of \ACRONYM{}. The sessions were transcribed, and a thematic analysis was performed by 2 members of the research team. While we uncovered many themes, some on implementation and interface design details, we focus on the broad conceptual challenges here. 

\textbf{Explanation specificity.} Our experts mentioned that users might find the current information provided by \ACRONYM{} rather vague. For example, when \ACRONYM{} identified potential disclosure of health related information, they anticipated that users would want to know why this would be of concern to them (e.g., higher insurance premiums). 
Similarly, in the case of fake news, our experts found the bot score indicators and credibility information insufficient to warrant a change in users' sharing habits. They expected that users would want to see what was specifically wrong about the bot (e.g., spreading of fake news or hate speech).

\textbf{Explanation depth.} \ACRONYM{} currently intercepts each post to provide information about potential unintended privacy disclosures individually.  However, institutional privacy risks often stem from analyzing historical social media data. Our experts expected \ACRONYM{} to provide explanations that help users understand how the current sharing instance would add to unintended inferences made from their posting history. They remarked that a combination of both depth and specificity was required to help users reflect meaningfully and that users should have the option to check the deeper explanations on demand.  

\textbf{Leveraging social and HCI theories.} Our experts also felt that for the \ACRONYM{} to be effective it had to be grounded in social theories. For example \ACRONYM{} could leverage various social theory-based interpretations of self- disclosure in tailoring effecting nudges or craft nudges which target specific cognitive biases to better help users understand the  privacy risks \cite{kokolakis2017privacy}. Similarly, in the example of preventing the sharing of fake news, the interface could incorporate explanations that appeal to social pressure or present posts that share an alternate point of view etc. \cite{10.1145/3290605.3300733}


\textbf{Designing for trust.} While \ACRONYM{} helps users through nudging, it in itself poses a risk due to its access to user data for generating nudges and explanations. Instead of deferring the question of trust to external oversight, legislation or practices such as open sourcing the design, experts felt that \ACRONYM{} needed to primarily perform all the inferences on the client side and borrow from progress in areas such as Secure Multiparty Computations~\cite{cramer2015secure}.

Summing up, the feedback from our experts regarding the current shortcomings of our prototype and future challenges closely match our proposed research agenda. A novel perspective stems from the consideration of social and HCI theories to complement the more technical research questions proposed in this paper.

%% file: sections/discussion.tex
\section{Discussion}
\label{sec:discussion}
This section covers (mainly non-technical) related research questions that are important but not part of our main research agenda.

\textbf{Ethical concerns.}
The line between nudging and manipulating can be very blurred. Efforts to guide users' behavior in a certain direction automatically raises ethical questions~\cite{Acquisti2017NudgesForPrivacyAndSecurity}. Nudging is motivated to be in the interest of users, but it is not obvious if the design decisions behind nudges and users' interests are always aligned. Even nudging in truly good faith may have negative consequences. For example, social media has been used to identify users with suicidal tendencies. Using privacy nudges to help users hide their emotional and psychological state would prevent such potentially life-saving efforts. Nudges might also have the opposite effects. 
Users might feel belittled by constant interventions such as warning messages. This, in turn, might make users more ``rebellious'' and actually increase their risky sharing behavior~\cite{Amin2020InfluencingPhotoSharing}. When, how often and how strongly to nudge are research questions need to be answered before the use of nudging in real-world platforms.

\textbf{Principle \& practical limitations.}
Algorithms for user nudging based on machine learning generally yield better results when more data is available. As such, social media platforms will likely always have the edge over solutions like \ACRONYM{}. 
Furthermore, a seamless integration of nudges in all kinds of environments is not straightforward. Our current browser extension-based approach is the most intuitive method. On mobile devices, a seamless integration would require standalone applications that mimic the features of official platforms apps, extended by user nudges. This approach is possible for platforms such as Facebook, Twitter or Instagram provide APIs that allow for the development of 3rd-party clients. ``Closed'' apps such as WhatsApp make a seamless integration impossible. Practical workarounds require the development of apps to which, e.g., WhatsApp messages can be sent for analysis.

\textbf{Nudging outside social media.}
In this paper, we focused on data-driven user nudging in social media. However, our in-situ approach using a browser extension makes \ACRONYM{} directly applicable to all online platforms. For example, we can inject the user nudges into any website, including Web search result pages, online newspapers, online forums, etc. However, such a more platform-agnostic solution poses additional challenges towards good UX/UI design to enable a helpful but also smooth user experience.

\textbf{Beyond user nudging.}
Automated user nudging is a promising approach to empower users to better face the threats on social media. However, user nudging is unlikely to be the ultimate solution but part of a wide range of existing and future efforts. From a holistic perspective of tackling social media threats, we argue that the underlying methods and algorithms required for effective user nudging -- particularly for risk assessment and visualization/explanation -- are of much broader value. Their outputs can inform policy and decision makers, support automated, human or hybrid fact-check efforts, improve privacy-preserving data publishing, and guide the definition of legal frameworks.

%% file: sections/conclusions.tex
\section{Conclusions}
\label{sec:conclusions}
This paper set out to accomplish two main tasks: outlining algorithmic threats when using social media, and proposing a research agenda to help users better understand and control threats on social media platforms through data-driven nudging. The fundamental goal of user nudging is to empower users to tackle the threats posed due to the information asymmetry between users and platforms providers or data holders. To scale and to compete with these algorithmic threats, user nudging must necessarily use algorithmic solutions, albeit, in an open and transparent manner. While this is multidisciplinary effort, we argue that the multimedia research community has to be a major driver towards leveling the playing field for the average social media user. To kick-start this endeavour, our research agenda formulates a series of research questions organized according to the main challenges and core tasks.
\\
\\
\textbf{Acknowledgements.} This research is supported by the National Research Foundation, Singapore under its Strategic Capability Research Centres Funding Initiative. Any opinions, findings and conclusions or recommendations expressed in this material are those of the author(s) and do not reflect the views of National Research Foundation, Singapore.